# Self-assembled contacts to nanoparticles using metallic colloidal spheres

C. R. KNUTSON[1], K. D. MCCARTHY[1], R. SHENHAR[2,3], V. M. ROTELLO[2], T. EMRICK[4], T. P. RUSSELL[4], M. T. TUOMINEN[1], A. D. DINSMORE[1]*
[1]Physics Department, University of Massachusetts, Amherst MA 01003, USA
[2]Chemistry Department, University of Massachusetts, Amherst MA 01003, USA
[3]Current address: Institute of Chemistry, The Hebrew University of Jerusalem, Israel 91904
[4]Polymer Science and Engineering Department, University of Massachusetts, Amherst MA 01003, USA

*e-mail: dinsmore@physics.umass.edu

**The spontaneous assembly of particles in suspension provides a strategy for inexpensive fabrication of devices with nanometer-scale control, such as single-electron transistors for memory or logic applications. A scaleable and robust method to form electrodes with the required nanometer-scale spacing, however, remains a major challenge. Here, we demonstrate a straightforward assembly approach in which metallic colloidal spheres serve as the electrodes. The devices are formed by assembly in suspension followed by deposition onto a patterned substrate. The key to this approach is that the inter-electrode (inter-sphere) spacing is spontaneously set to allow tunneling contact with a single layer of nanoparticles. The measured current exhibits the Coulomb blockade owing to the small size and large electrostatic charging energy of the nanoparticles. We show that the device resistance can be tuned by means of a gate electrode. Our results demonstrate an altogether new approach to inexpensive and large-scale fabrication of electronic devices such as transistors with nanometer-scale features.**

Assembly of particles in a liquid suspension is an attractive route to materials fabrication owing to the potential for high spatial resolution, unusual properties, and low cost. For example, nanoparticles may be placed on or between electrodes on a patterned substrate to form single-electron transistors[1-4] for memory or logic applications[3-6]. This approach, however, still requires that electrodes be formed using a high-resolution method such as break-junction[7,8], controlled etching[9,10], electromigration[11-14], or lithography followed by controlled deposition[15-18]. Here we demonstrate a highly pragmatic new approach to forming electronic materials and devices, with the spacing between electrodes controlled via self-assembly to allow tunneling contact. As a demonstration, we have fashioned transistors in which colloidal metallic spheres form tunneling contacts with ligand-stabilized nanoparticles. Our approach allows the formation of a large number of devices in suspension, which are then deposited on a substrate with leads that could be formed with microcontact printing or other large-area methods[19,20].

We used micron-scale spheres composed of a solid or molten metal, which were suspended in oil and coated with a layer of colloidal gold nanoparticles[21-23]. Figure 1 shows 120-μm-diameter Woods Metal (WM) spheres coated with gold nanoparticles and then deposited on a substrate with conducting leads (details provided in the Methods section). After deposition, the coated spheres adopted a spacing comparable to the nanoparticle size owing to attractive capillary[24], electrostatic, and van der Waals[25] forces. Thus, the current flowing from one lead to the other passed through one or more junctions that contained a layer of nanoparticles (inset of Fig. 1). For the electronic devices presented here, we used gold nanoparticles stabilized by undecanethiol ligands[26]; the mean nanoparticle diameter was 1.7 nm with a standard deviation of 0.3 nm and an approximately log-normal size distribution. In a separate experiment, we mixed the WM spheres with 3.2-nm-diameter CdSe nanoparticles stabilized by tri-*n*-octylphosphine oxide (TOPO), then verified with fluorescence microscopy that the nanoparticles had adsorbed on the metal surfaces. Adsorption of the nanoparticles is driven by the WM droplets' surface tension[21,27] and by van der Waals attraction between the nanoparticles and the metal surfaces.

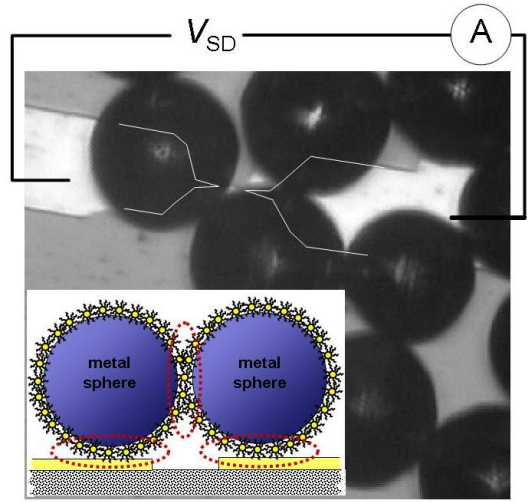

**Figure 1 Self-assembled device with Woods Metal spheres coated with Au nanoparticles.** The 120-μm WM spheres were deposited from toluene onto an oxidized Si substrate with patterned Au electrodes (shown in white, with outlines added). The sample was exposed to air, allowing the toluene to evaporate. (Brightfield optical microscope image.) *Inset,* Schematic side-view of the device. The dashed ovals highlight the three contacts among the nanoparticle-coated WM and the substrate (not to scale).

Figure 2 shows the current ($I$) through the device shown in Fig. 1, measured as a function of applied voltage ($V_{SD}$). The data in Fig. 2a were acquired soon after deposition of the WM/nanoparticle spheres. Initially, the response was ohmic with a low resistance ($R = 15.5$ Ω), owing to direct contact between the asperities on the WM surfaces that appeared during cooling and freezing of the molten droplets[28]. On increasing $V_{SD}$ to 1.15 V, the resistance dropped discontinuously by a factor of approximately 1.4. Upon decreasing the bias voltage, the device remained ohmic until the current dropped at $V_{SD} \approx$ 1.3V, after which it followed the linear plus cubic form that is typical of tunneling (inset of Fig. 2a). This transition from ohmic to tunneling behavior was observed in other devices as well, including those composed only of WM and oil (*i.e.* without nanoparticles or added surfactants).

Immediately after the first $I$-$V_{SD}$ trace, the device showed a further reduction in current and clear evidence of the Coulomb blockade. The four subsequent $I$-$V_{SD}$ scans (Fig. 2b) display variations, but the overall behavior is consistent: $I$ was markedly suppressed when $V_{SD} < 0.3$V, and increased approximately linearly at larger $V_{SD}$. This trend is consistent with the Coulomb blockade model, where the current is suppressed until $V_{SD}$ approaches the potential needed to overcome the electrostatic energy of placing an electron on a nanoparticle[4,29]. The orthodox Coulomb blockade theory applied to metallic particles of 1.4-nm diameter with 0.7-nm organic ligands (serving as tunnel barriers) predicts a gap of approximately $e/C \approx$ 0.2-0.3 V at zero temperature, where $C$ is the total capacitance of an individual nanoparticle and is estimated by approximating the electrodes as parallel plates[4]. Previous experiments with ligand-stabilized colloidal Au nanoparticles in the 1-3-nm range revealed threshold voltages in the range of 0.1-1V[14,30]. These values are in good agreement with our data, indicating the presence of a monolayer of Au nanoparticles in one or more of the junctions. Modeling the nanoparticles as independent parallel resistors with $R \sim 10^9$ Ω, we estimated from the differential resistance above threshold that there were of order $5\times10^3$ nanoparticles conducting current. Similar results were obtained for the current between nanoparticle-coated WM spheres in contact with 100-μm Pt wires in solution, though this device architecture was less mechanically stable. Control experiments on devices without nanoparticles did not display this current suppression.

Cooling the devices to 77 K by immersion in liquid nitrogen shows the Coulomb blockade more strikingly (Fig. 2c). Here, the threshold behavior was more pronounced than at room temperature because of the reduced contribution from thermally excited transport over the electrostatic-energy barrier. The threshold voltage was approximately 0.6 V, which is larger than in the room-temperature scans. As a result of the mechanical stress from plunging the sample into liquid $N_2$, more than one of the three junctions in series may have been of a high-resistance tunneling nature, so that the voltage drop at a single junction was less than $V_{SD}$.

The features of the current-voltage trace are readily explained by melting of the WM surface near the contact regions owing to resistive heating. We estimated the temperature near the contacts assuming that the power ($I \times V_{SD}$) was dissipated inside a volume of $(100nm)^3$ and that the heat was conducted through the WM without convection. At $V_{SD}$ = 1.15 V, this led to an estimated temperature far above the WM melting temperature of 73-77 °C. We conclude that the asperities on the WM surfaces melted, thereby increasing the contact area and reducing the resistance. Similarly, this model suggests that the current fluctuations at $V_{SD} > 1.6$ V arose from boiling of the toluene (boiling point ~ 110 °C) in the layer between the particles. The precipitous drop in current upon decreasing $V_{SD}$ likely arose from re-freezing of the WM near the junction; since WM expands on freezing[31], this may have disrupted the gap and allowed toluene to enter and wet the exposed WM surfaces. Separate experiments showed that toluene wets the WM spheres.

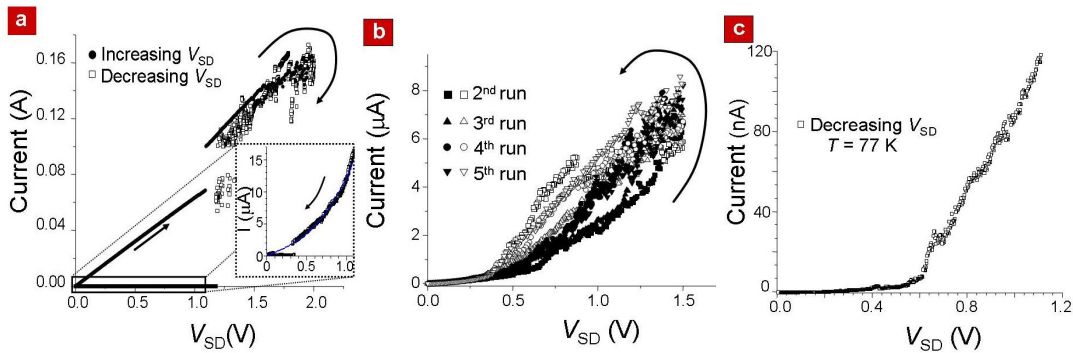

**Figure 2 Measured current *vs.* applied voltage ($V_{SD}$) showing the development of the Coulomb blockade. a**, The first voltage cycle after deposition is shown for increasing (●) and decreasing $V_{SD}$ (□) at room temperature. **inset,** Data for decreasing $V_{SD}$, in expanded scale. The dashed curve shows a linear+cubic fit to the current; the agreement suggests tunneling behavior. **b**, The four subsequent scans, showing the Coulomb blockade and hysteresis. **c**, $I$-$V_{SD}$ scan after cooling the device to 77 K by immersion in liquid nitrogen. The threshold near 0.6 V arose from the Coulomb blockade.

The formation of a layer of nanoparticles and the Coulomb blockade are readily explained by migration of the nanoparticles into the gap. Owing to the applied voltage, the nanoparticles must experience a dielectrophoretic force toward the gap, which arises from the interaction of the dipole induced in each nanoparticle with the gradient in the electric field near the junction. Approximating the geometry near the junction as the contact between two spheres, we estimated that the force on a nanoparticle 100 nm away from the gap was of order 10 – 1,000 fN. In the presence of such a force, nanoparticles in the viscous interfacial layer move toward the gap at a speed given by the force divided by the friction coefficient (which depends on the thickness and viscosity of the thin interfacial layer of toluene, nanoparticles, and ligands). After a sufficient density of nanoparticles accumulates in the junction, direct tunneling through the solvent becomes negligible, tunneling through the nanoparticle ligands dominates the transport, and the Coulomb blockade can arise. Similarly, the mobility of the nanoparticles explains the hysteresis seen in Fig. 2b: as $V_{SD}$ increases, the field draws more nanoparticles into the gap and the current is enhanced. After $V_{SD}$ returns to zero, osmotic pressure within the nanoparticle layer gradually restores the nanoparticle density to its steady, zero-voltage state.

As further evidence of the gold-nanoparticle layer and the Coulomb blockade, we changed the current through a device by applying voltage to a gate electrode. Here, solid Al particles were suspended in toluene with the gold nanoparticles, then rinsed and deposited on a patterned substrate (details provided in the Methods section). The initial $I$-$V_{SD}$ scan showed high resistance (~ $10^{15}$ Ω) with $V_g = 0$. After increasing $V_g$ to 1400 V for 5 s then returning to $V_g = 0$, however, the Al particle was pulled into contact with the leads by dielectrophoretic attraction between the Al particle and the gate. Following this step, we obtained repeatable and non-hysteretic $I$-$V_{SD}$ plots, which show the threshold behavior seen in WM devices (Fig. 3). The measured differential conductance ($dI/dV_{SD}$) initially increased with $V_{SD}$, then peaked at $V_{SD} \approx$ 0.6 V (upper inset). The conductance at $V_{SD} = 0$ increased with $V_g$. The conductance at $V_{SD}$ = 0.6 V, however, decreased from 14 to 11 (MΩ)$^{-1}$ (lower inset). This trend cannot readily be explained either by tunneling directly between the Al particle and the electrodes or by leakage of current from the gate. Rather, the drop in conductance arose from an effective charge induced on the nanoparticles by the gate voltage ($V_g$ × the gate capacitance). In the orthodox Coulomb blockade theory, this induced charge shifts the threshold voltage away from its zero-gate value and changes the conductance. We numerically computed the conductivity *vs.* $V_g$ at $T$ = 300 K using the orthodox theory with $C$ = 0.25×10$^{-18}$ F and a gate capacitance of 5×10$^{-23}$ F; the same trend of d$I$/d$V_{SD}$ was found. The polydispersity of nanoparticle sizes and a possible random distribution of static charges, however, prevent quantitative agreement with the data. Nonetheless, the gate effect of this transistor provides strong evidence of the Coulomb blockade and tunneling contact with a monolayer of nanoparticles in the gap. The value of $V_g$ required for switching can be greatly reduced by increasing the gate capacitance, *e.g.*, by using smaller metal spheres.

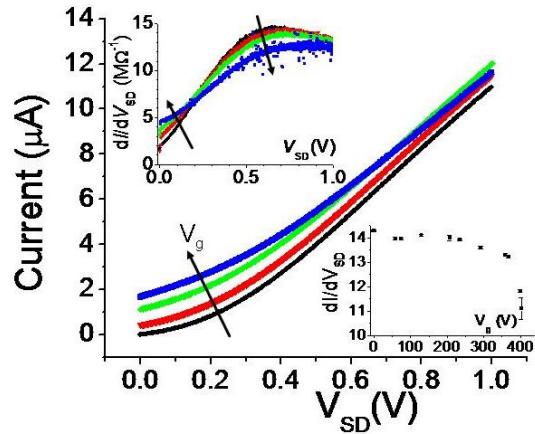

**Figure 3 Measured current *I* vs. *V*$_{SD}$ for a device with a gate electrode**. The gate voltages are $V_g = 0$ (●), 207 (▼), 312 (♦) and 400 V (■). The arrows show the trend with increasing $V_g$. This device was formed from a single solid Al sphere straddling two patterned electrodes at room temperature. The small but non-zero current at $V_{SD} = 0$ and $V_g > 0$ arose from leakage of current from the gate to the drain electrode. **upper inset**, Differential conductance $dI/dV_{SD}$ *vs.* $V_{SD}$, in units of (MΩ)$^{-1}$. **lower inset,** $dI/dV_{SD}$ *vs.* $V_g$ at $V_{SD} = 0.6$ V. The change in conductance shows the gate effect.

Our results demonstrate the use of metal microspheres as self-assembling electrodes for tunneling devices such as transistors. We have also demonstrated a range of new phenomena, including the transition from ohmic to tunneling behavior, the Coulomb blockade, and the gate effect. This method can be combined with other techniques, such as the use of electrohydrodynamic[32] or capillary[24,33] forces, to arrange spheres precisely at chosen sites on a substrate. Moreover, because the junction spacing is formed spontaneously, a broken junction can be repaired. This approach to forming working devices offers other advantages owing to the simplicity of the method and the potential for extremely low cost and large-area coverage.

## METHODS

Monodisperse metal droplets were formed by emulsification of Woods Metal (WM, an alloy of 50% Bi, 25% Pb, 12.5% Sn, 12.5% Cd; $T_{melt}$ = 73-77 °C)[34]. The molten metal was forced through a micropipette into a spinning cup containing oil (Castor oil or hexadecane) without surfactants[28]. The flowing oil broke the droplets at a consistent size[35] and a macroscopic quantity of droplets was obtained. At room temperature, the WM droplets solidified and were stable against coalescence, though the surfaces were rough[28]. Solid Al spheres were received as a powder (Aldrich). The metal spheres (Al or WM) were subsequently suspended in toluene containing nanoparticles, then washed in toluene to remove excess nanoparticles. Droplets of the suspension were placed on the patterned substrate and exposed to the air to allow the toluene to evaporate.

For the WM-based device, the gold electrodes were as shown in Fig. 1. For the Al-based device (corresponding to Fig. 3), two parallel gold electrodes, separated by a 10 μm wide gap, were straddled by a single 30-μm solid-Al particle. The doped silicon substrate, which was buried beneath a 100-nm thick oxide layer, served as the gate. The Al-particle devices were more mechanically stable than the WM devices in the presence of the large gate field.

The gold electrodes were connected to a DC power supply in series with a current amplifier (DL Instruments #1211)[28]. A bias voltage $V_{SD}$ was applied between the two electrodes and the current $I$ through the device was measured. In a typical scan, $V_{SD}$ was ramped from zero to 1 or 2 V, then back to zero during a time interval of approx. 15 min. The device was sealed inside a metal box to suppress noise.

## Acknowledgements

ADD and CRK gratefully acknowledge support from the Research Corporation Cottrell Scholarship Program and a UMass Faculty Research Grant. We also acknowledge support from the NSF-supported Materials Research Science and Engineering Center on Polymers (DME-0213695) as well as NSF grants DMR-0306951, DMI-0103024 and CHE 0518487.